\documentclass[12pt]{article}

\usepackage{bbold}
\usepackage{amssymb}
\usepackage{graphics}
\usepackage{epsfig}
\def\N{{\cal N}}

\def\s{\sigma}

\def\wt#1{\widetilde{#1}}

\def\m{\mu}
\def\n{\nu}

\def\be{\begin{equation}}
\def\ee{\end{equation}}
\def\beq{\begin{equation}}
\def\eeq{\end{equation}}
\def\bea{\begin{eqnarray}}
\def\eea{\end{eqnarray}} 
\def\beqa{\begin{equation}\begin{array}{l}}
\def\eeqa{\end{array}\end{equation}}
\def\eqn#1{(\ref{#1})}
\def\eqref#1{eq.~(\ref{eq:#1})}

\def\a{\alpha}
  \def\g{\gamma}

\def\nn{\nonumber}

\def\tr{{\bf tr}}
\def\div{{\bf div}}
\def\grad{{\bf grad}}
\def\N{{\bf N}}
\def\g{{\bf g}}

\begin{document}

\thispagestyle{empty}
%\begin{flushright}
%\framebox{\small BRX-TH~548
%}\\
%\end{flushright}

\vspace{.8cm}
\setcounter{footnote}{0}
\begin{center}
{\Large{\bf 
Partially Massless Spin 2 Electrodynamics
}
    }\\[10mm]

{\sc S. Deser$^\sharp$
and A. Waldron$^\natural$
\\[6mm]}

{\em\small  
$^\sharp$Lauritsen Lab,
Caltech,
Pasadena CA 91125
 and Physics Department, Brandeis University, Waltham,
MA 02454, 
USA\\ {\tt deser@brandeis.edu}}\\[5mm]
{\em\small  
$^\natural$Department of Mathematics,
University of California, Davis, CA 95616,
USA\\ {\tt wally@math.ucdavis.edu}}\\[5mm]

\bigskip

\bigskip

{\sc Abstract}\\
\end{center}

{\small
\begin{quote}

We propose that maximal depth, partially massless,
higher spin excitations can mediate charged matter
interactions in a de Sitter universe. The proposal is 
motivated by similarities between these theories
and their traditional Maxwell counterpart: their
propagation is lightlike and corresponds to the same Laplacian
eigenmodes as the de Sitter photon; they are conformal in
four dimensions; their gauge invariance has a single scalar parameter
and actions can be expressed as squares of single derivative curvature tensors.
We examine this proposal in detail for its simplest spin~2 example.
We find that it is possible to construct a natural and consistent interaction scheme to conserved 
vector electromagnetic currents primarily coupled to the helicity~1 partially massless modes. 
The resulting current-current single ``partial-photon'' exchange amplitude is the (very unCoulombic) sum of contact and shorter-range terms, so the partial photon cannot replace the traditional one,
but rather modifies short range electromagnetic interactions. 
We also write the gauge invariant fourth-derivative effective actions that might appear as effective corrections to the model, and their contributions to the tree amplitude are also obtained.

\bigskip

{\tt PACS: 03.70.+k, 04.62.+v, 11/15.-q}

\bigskip

\end{quote}
}

\newpage
%\setcounter{page}{1}
%%%%%%%%%%%%%%%%%%%%%%%%%%%%%%%%%%%%%%%%%%%%%%%%%%%%%%%%%%%%%%%%%

%\vfill

%\tableofcontents

%\vfill

%\newpage

\section{Introduction}

Some time ago~\cite{Deser:2001pe}, we developed a series of ``partially massless'' higher spin theories in (Anti) de Sitter
((A)dS) space, characterized by higher derivative invariances under lower rank gauge parameters than their strictly massless counterparts. These models generalized the lowest, spin~2, example~\cite{Deser:1983tm}
 and were seen to have gratifying properties, such as light cone propagation~\cite{Deser:2001xr}, locally positive energy~\cite{Deser:2001wx}, they irreducibly represent the dS isometry group unitarily~\cite{Deser:2003gw}  and  possess a clear hierarchy of ghost-free helicity excitations ranging from~$\pm s$ as far down 
as~$\pm t$
at depth $0<t<s$ (with $s-t$ equaling the gauge parameter rank).
In this paper we concentrate on maximal depth $t=s-1$ theories which 
have a scalar gauge parameter and the dS Maxwell model
is the first member of this series of theories. Moreover, in dimension four, all these theories
are distinguished by being conformally invariant~\cite{Deser:2004ji}. They all  propagate in the 
same 
way save for the additional helicities, and are describable in terms 
of curvature tensors first order in derivatives. The purpose of this paper is to investigate
whether this intriguing string of coincidences has a deeper physical significance. Namely whether
maximal depth partially massless theories can mediate dS electromagnetic interactions.

Initially, we did not consider possible interactions of these systems with conventional matter sources, a gap we attempt to fill here. For concreteness, we concentrate on the lowest, spin 2, model, represented by a symmetric tensor field $A_{\mu\nu}$
whose ``natural'' source would of course be the stress tensor, but that has been preempted by another, massless, spin 2 field.
The (unique) partially massive model here involves both $(\pm2, \pm1)$ helicities and a scalar gauge parameter. We will therefore attempt to interpret these excitations as ``partial photons'' and
focus on (conserved) vector current matter sources that primarily excite helicity~1, as a sort of pseudo-electrodynamics. 

Our preliminary investigations have uncovered a consistent coupling to charged matter for
spin~2 partial photons. An analysis of one-particle exchange amplitudes yields a sum of
short range and contact charged matter interaction which indicates that the traditional dS photon
cannot be replaced, but rather only supplemented by its partial counterpart. We also 
present a study of higher derivative effective actions to exhibit possible
radiative corrections to the leading order tree level analysis.

In the next Section, we outline the properties of our model and recast it in Maxwellian form in terms of scalar-gauge invariant first-derivative field strengths, rather than the Riemann-like curvatures associated with the massless tensor. In Section~\ref{couplings} we introduce sources and in its following Section elaborate on the resultant current-current interactions.
Finally we exhibit the form of quartic derivative corrections to the original free field model, and their effects on these couplings. An Appendix summarizes the symmetric algebra formalism 
of~\cite{Hallowell:2005np} which makes our detailed computations possible.

\section{Partially Massless Spin~2: Partial Photons}
\label{HS}
Our dS conventions\footnote{Actually, all computations in this paper apply also to AdS backgrounds, but partially massless excitations are no longer unitary in that case.} with cosmological constant $\Lambda>0$ are
\be
R_{\m\n}{}^{\rho\s}=-\frac{2\Lambda}{3}\
\delta^{\rho}_{[\m}\delta^{\s}_{\n]}\, ,
\ee
and the commutator of covariant derivatives acting on vectors is
\be
[D_\mu,D_\nu]V_\rho=\frac{2\Lambda}{3}\ g_{\rho[\mu}V_{\nu]}\, .
\ee
The dS metric $g_{\m\n}$ moves all indices and defines covariant derivatives; its signature is $(-+++)$ in four dimensions.  
Throughout our analysis,  we hold this dS background fixed.                                                                                 

Unlike the action and field equations of its strictly massless  de Sitter graviton relative,
the partially massless spin~ 2 excitations (``partial photons'') 
can be formulated in terms of a Maxwell-like
curvature tensor that is first order in derivatives\footnote{In fact, the same holds
for maximal depth theories of arbitrary spin~\cite{Gover}.}
\be
F_{\mu\nu\rho}=D_\mu A_{\nu\rho}-D_\nu A_{\mu\rho}\, ,
\ee
where the potential $A_{\mu\nu}=A_{\nu\mu}$. The curvature $F_{\mu\nu\rho}$ is 
invariant under gauge transformations
\be
\delta A_{\mu\nu}=\Big(D_\mu D_\nu +\frac\Lambda 3 g_{\mu\nu}\Big)\, \alpha\, .\label{gauge}
\ee
Although one would usually expect curvature tensors for spin~2 fields $A_{\mu\nu}$ 
to  be of Riemann type -- second order in 
derivatives~\cite{deWit:1979pe}, the additional derivative in the 
gauge transformation balances the one ``missing'' in the curvature.
Moreover, the gauge parameter $\alpha$ is a scalar, just like the
Maxwell case. This motivates our main observation that the partially massless
spin~2 field may be better viewed as a generalization of the photon, rather than 
its graviton antecedent.

On the basis of the gauge invariance~\eqn{gauge} alone, there exists a one parameter family
of invariant actions and accompanying field equations. However, requiring the absence of
ghosts in the free spectrum yields a unique ``partial photon'' action principle\footnote{To avoid confusion, note that this is simply a rewriting of the usual, second order, $s=2$, partially massless action obtained by linearizing the Einstein tensor. Observe also that there exists a natural first order reformulation of the above action.}
\be
S_{\rm pp}=-\frac14 \int d^4x \sqrt{-g}\Big[F_{\mu\nu\rho}F^{\mu\nu\rho}+F_\mu F^\mu\Big]\, ,\label{Spp}
\ee
where $F_\mu$ is the curvature trace
\be
F_{\mu}\equiv F_{\mu\nu}{}^\nu\, .
\ee
To be precise, unitary, spin~2 irreducible representations of the dS isometry group $SO(4,1)$
carry either 2, 4 or 5 degrees of freedom (respectively strictly massless, partially massless
or massive theories)~\cite{Deser:2003gw}. Here the ten covariant field components $A_{\mu\nu}$
yield 4 partially massless  degrees of freedom because the free field equations
\be
{\cal G}_{\nu\rho}=D^\mu F_{\mu\nu\rho}+\frac12 g_{\nu\rho}D^\mu F_\mu
-\frac12 D_{(\nu} F_{\rho)}=0\, ,\label{eom}
\ee 
obey the constraint
\be
D^\nu {\cal G}_{\nu\rho}=\frac{2\Lambda}{3}\ F_\rho\, .\label{constraint}
\ee
This removes four degrees of freedom and two more are accounted for by the gauge
invariance~\eqn{gauge} and corresponding Bianchi identity
\be
D^{\nu}D^{\rho}{\cal G}_{\nu\rho}+\frac{\Lambda}{3}\  {\cal G}^\nu{}_\nu\equiv 0\, ,
\ee
leaving four physical propagating modes. These correspond
to excitations of helicity $(\pm2, \pm 1)$. They propagate at the speed of light~\cite{Deser:2001xr}
({\it i.e.} along the light cone, dS being conformally flat) and obey a local energy
positivity theorem~\cite{Deser:2001wx} completely analogous to dS gravitons~\cite{Abbott:1981ff}.
Moreover, just like photons in four dimensions (but in contrast to gravitons), 
the partially massless spin~2 theory is conformally invariant~\cite{Deser:1983tm,Deser:2004ji}.

We are now ready to introduce sources.

\section{Couplings}

\label{couplings}

As stated, the stress tensor being the source of gravity, we turn to the other universal possibility, the covariantly conserved vector current $J_\mu$, 
\be
D_\mu J^\mu=0\, .
\ee 
The coupling obviously requires an extra index, so there are two possible local combinations,
\be
S_{\rm int} = -\int d^4x \sqrt{-g}  A_{\m\n} ( Q D^\m J^\n + Q' g^{\m\n} D_\rho J^\rho)\, .   \label{co}                                                                       
\ee
The charges $Q$ and $Q'$  carry mass dimension unity.
[For conserved $J^\mu$, the $\tr A\  \div J$ term is moot at tree level but can play a r\^ole in 
loops which we mostly ignore in this work\footnote{Observe also that the tensor $D_{(\m} J_{\n)}$ is
trace-free onshell. This does not mean that the coupling is conformal, because this tensor ought
not be confused with its stress energy counterpart. }.]  A quick  computation reveals
\bea
\delta \Big(D^\nu A_{\mu\nu}\Big)&=&D_\mu \Big\{\Big(D^2+\frac{4\Lambda}3\Big)\alpha\Big\}\, ,\nn\\[1mm]
\delta \Big(D_\mu A_\nu{}^\nu\Big)&=&D_\mu \Big\{\Big(D^2+\frac{\Lambda}{3}\Big)\alpha\Big\}\, .\label{phg}
\eea
Hence the combination
\be
V_\mu=Q D^\nu A_{\mu\nu} + Q' D_\mu A_\nu{}^\nu
\ee 
transforms like an electromagnetic potential $\delta V_\mu=D_\mu \wt \alpha$ with
parameter $\wt \alpha=
(Q+Q') D^2\alpha + \frac{Q+4Q'}{3}\Lambda \alpha$. Hence we can hope to couple it
consistently to charged matter fields.
There are two distinct cases:
\begin{enumerate}
\item {\it Non-dynamical matter:} For an on-shell, conserved, background
matter current, the $Q'$ coupling is irrelevant and as $\div A$
transforms into the gradient of a scalar (just like an E/M potential -- see~\eqn{phg}),
the interaction $S_{\rm int}$ preserves gauge invariance.
\item {\it Dynamical matter:} The interacting partial photon -- charged matter system
$$ S= S_{\rm pp} + S_{\rm int} + S_{\rm matter}\, , $$ 
varies under partial gauge transformations with parameter $\alpha$ and local $U(1)$ 
transformations with parameter $\beta$ as
$$
\delta S= \int d^4 x \sqrt{-g} J^\mu D_\mu \Big\{
-e \beta + (Q+Q') D^2 \alpha + \frac{4Q+Q'}{3}\ \Lambda \alpha\Big\}\, .
$$
For the choice of parameters $$Q=-Q'\, ,$$ we may identify 
$Q\Lambda\alpha=\beta$
and the system is invariant under arbitrary $U(1)$ gauge transformations. For general
choices of $Q$ and $Q'$ we still have invariance under arbitrary partial gauge transformations 
$\alpha$, so the system is consistent\footnote{The key point being the ability to  
gauge away ghostlike partial photon excitations.}, but there exist (a set of measure zero) $U(1)$
gauge equivalent matter configurations not reachable by any choice of~$\alpha$,
corresponding to zero modes of the operator $(Q+Q') D^2 + \frac{4Q+Q'}{3}\ \Lambda$.
\end{enumerate}
For the remainder of our analysis we retain both parameters $(Q,Q')$ and the distinction
between dynamical and background charged matter will not play any special r\^ole.

Classical consistency of the coupling~\eqn{co} relies not only on the gauge invariance~\eqn{gauge}
but also the constraint~\eqn{constraint} to ensure that ghost states are non-propagating. In particular
one might worry that including the source $J^\mu$ introduces terms involving covariant derivatives
of dynamical fields to the right hand side of~\eqn{constraint}. In particular, the key property that the constraint
is only first order in time derivatives of fields\footnote{In this analysis time can be taken
as the coordinate along any timelike vector field. In dS, there is a timelike Killing
vector (within the intrinsic horizon) which provides the most natural choice of
time-slicing.}, could be violated. In fact, there is actually no obstruction to the constraint
analysis, because the new contributions only involve matter fields. In the case that these
are dynamical, unwanted time derivatives can always be removed using the matter
field equations.

We  next study the simplest phenomenological implications of our new coupling, one particle exchange processes.                              

\section{One Particle Exchange}

\begin{figure}
\begin{center}
\epsfig{file=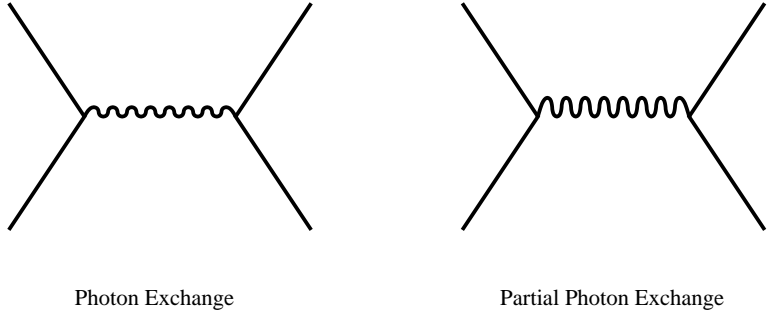,width=4in}
\end{center}
\caption{Partial Photon Mediated Scattering. \label{graph}}
\end{figure}

The simplest effect of the coupling~\eqn{co} is clearly the ``one-partial photon'' exchange process
depicted in Figure~\ref{graph}.
There are two distinct phenomenological possibilities: We could replace the photon by its partial counterpart and compute 
only the second diagram -- an option quickly ruled out by the results that follow-- or view the partial photon as a modification               
of the existing electromagnetic theory and study the sum of photon and partial photon exchange diagrams.
The calculation begins with the propagator for massive spin~2 fields~\cite{Lichnerowicz}
\bea
{\cal D}&=&\frac{1}{\square-m^2+6}\left[
1-\frac{\grad\ \div}{m^2}\right. \nn\\[2mm]&+&\left.\frac{ \grad^2\div^2
-\frac12 m^2(m^2-3) \g\ \tr +\frac12 m^2(\g\ \div^2+\grad^2\tr)}{3m^2(m^2-2)}
\right]\, .\nn\\
\eea
Here we employ units $\Lambda/3=1$ and the operator $\square$ is Lichnerowicz's wave operator~\cite{Lichnerowicz}. The 
operators $(\tr,\div,\grad,\g)$ correspond to the trace, divergence, gradient and multiplication by the metric and symmetrizing operations
in the symmetric algebra formalism of~\cite{Hallowell:2005np}. A self contained account is given in Appendix~\ref{symm}.
The first physical observation is that there are poles for masses $m^2=0$ and $m^2=2=2\Lambda/3$. These are easily understood
as corresponding to the strictly massless graviton and partially massless limits where gauge invariances imply non-invertibility
of the kinetic term. In this connection, the coefficient $(m^2-3)/(m^2-2)=(m^2-\Lambda)/(m^2-\frac{2\Lambda}3)$ of $\g\ \tr$ is also interesting, because it
is the basis of the resolution to the Veltman-van Dam--Zhakarov ambiguity~\cite{vanDam:1970vg,Zakharov:1970cc}. 
Namely, when sandwiched between covariantly conserved 
stress tensors (so all terms involving $\grad$ or $\div$ vanish), the spin~2 propagator limits to its massless flat space counterpart
when one first takes the mass $m^2\rightarrow 0$ and thereafter considers vanishing cosmological constant $\Lambda\rightarrow 0$~\cite{Kogan:2000uy,Porrati:2000cp,Deser:2000de}.

The partially massless limit $m^2\rightarrow 2$ can be taken safely in the exchange amplitude 
\be
{\cal A} = \int T {\cal D} T
\ee
for (two index-symmetric) sources $T\equiv T_{\mu\nu}dx^\mu dx^\nu$ obeying the partially massless conservation law
\be
(\div^2+\tr)T=0\, .
\ee
Explicitly we find
\be
{\cal A}=\int T\ \frac{1}{\square+4}\left[1-\frac12\grad\ \div -\frac{\square+\frac83}{4\square}\g\ \tr\right] T\, .
\ee
Specializing to the coupling~\eqn{co}, which onshell yields
\be
T=Q\  \grad J\, ,
\ee
produces the exchange amplitude
\be
{\cal A}_{\rm pp}=\frac{Q^2}2\int J (\square + 6) J\, .\label{App}
\ee
Clearly, by themselves, this sum of contact and short range terms 
is a rather unphysical amplitude for the interaction of charged matter.
If we include the first, photon, diagram of figure~\eqn{graph} and call the dimensionless
quantity (reinstating $\Lambda$)
\be
q^2\equiv \frac{Q^2\Lambda}3\, ,
\ee
we find 
\be
{\cal A}_{\rm tot}=\int J \ \frac{e^2}{\square}\left\{1+\frac{q^2}{2e^2}\frac{3\square}{\Lambda}\Big(\frac{3\square}{\Lambda}+6\Big)\right\}J\, .\label{AEM}
\ee
Thanks to the powers of inverse $\Lambda$, only extremely large values of the coupling~$q$ will produce measurable effects from the new  contact and short range interactions
predicted by this result.

Finally,  lest the reader be disappointed by the impossibility of replacing Maxwell
photons by their dS partial spin~2 counterparts, as expressed by the amplitude ${\cal A}_{\rm pp}$
in~\eqn{App}, we provide a simple rederivation of our results relevant for the computations of 
candidate radiative counterterms studied in the next Section: 

The one-particle exchange amplitude computation requires us to compute
$\int (\grad J) A(J)$ with $A=A(J)$ determined by
\be
GA=\grad J\, ,\label{eqn}
\ee
where $G A$ are the partially massless field equations and the kinetic operator~$G$ is as in~\eqn{G}.
Since the current $J$ is a transverse vector ($\div J=0$), we need only compute the helicity~1
part of $A=\ldots + \grad A^T +\ldots$ where $A^T$ is also a transverse vector.
Using the constraint~\eqn{Gcon} to compute the divergence of~\eqn{eqn}
yields
\be
-2\div \ \grad A^T=\div \ \grad J\, .
\ee
The $[\div,\grad]$ commutator gives an overall  factor $\square + 6$ so $A^T=-\frac12 J$
and in turn $\int (\grad J) A(J)=\frac12 \int J \div\ \grad J = \frac 12 \int J(\square+6)J$
as claimed. Notice the crucial r\^ole played by the divergence
constraint in this derivation. 

\section{Higher Derivative Actions}

It is unlikely that the form of the amplitude~\eqn{AEM} is respected by radiative corrections.
Alternatively one might like to search for modifications of our underlying theory in order to
produce ``improved'' amplitudes. From either viewpoint, an interesting question is
what higher derivative corrections are allowed to the result~\eqn{AEM}. While we avoid, 
at this premature stage, detailed higher loop computations, much can be said about candidate local     counterterms. We begin with the most rudimentary requirement, namely that they be invariant under the 
partial gauge transformation~\eqn{gauge}.  Moreover, we restrict our attention
to corrections quartic in derivatives and again employ the symmetric algebra formalism. 
In particular we specify actions by displaying the analog of the ``kinetic operator'' $G$ in~\eqn{G},
in terms of which the field equations are $G A=0$ and action $S=-\frac 12\int A G A$ (which is equivalent to~\eqn{Spp}).
One-particle exchange amplitudes are obtained by solving the analog of $G A=\grad J$.

The most general partially gauge invariant action quartic in derivatives~is\footnote{It is interesting
to note that higher derivative Maxwell-like actions have also been considered
in a mathematical context, where one attempts to preserve conformal invariance
in dimensions higher than four~\cite{Branson}.}
\bea
S_4&=&(\alpha_1-1) G  +(\alpha_2-1)\grad (\square-\grad\ \div) \div \nn\\[4mm]
       &+& \alpha_3 ( \g\square\tr - \g \ \div^2-\grad^2\ \tr  + \grad\  \div)\nn\\[4mm]
       &+&\beta_1 ([\square+6][\square+4]-4\grad\ \div-\frac32\  \g[\square+2]\tr\nn\\ &&\quad
                              +\g\ \div^2+\grad^2\ \tr-\grad[\square-\frac12\grad\ \div]\div)\nn\\[4mm]
       &+&\beta_2(\g\ \square^2\tr -\g\square\div^2-\grad^2\square\ \tr+\grad^2\div^2)\, .
\eea
The total action for the intermediate partial photons is 
\be
S_{\rm tot}=G+S_4\, .
\ee       
This theory produces the amplitude
\be
{\cal A}_{\rm tot}=\int J\ \frac1{(1-\alpha_2)\square-\alpha_3+\frac{2\alpha_1}{\square+6}}\ J\, ,
\label{qAMP}
\ee
about which we observe the following:                                          
\begin{enumerate}
\item At $\a_1=\a_2=\a_3$ we obtain a photon amplitude $J\frac1\square J$.
However, although this counterterm is by construction gauge invariant, to 
avoid propagating ghosts, one might also try to impose the divergence constraint
on the counterterm. Yet for this choice of parameters we find $\div S_4=\grad\square\div^2+\cdots$
which certainly violates the constraint.
\item At  $\alpha_1= \alpha_2=1$, $\alpha_3=0$, $\beta_2=-\beta_1/6$, one obtains
$\div S_4=0$, and hence leaves the divergence constraint unaltered. However 
this case returns to the original amplitude~\eqn{App}. In fact, this conclusion is obvious, 
since only the helicity~1 part of $S_4$ can contribute to the exchange amplitude.
\item One can also consider intermediate situations where (i) the leading derivative contributions to
$\div S_4$ vanish or (ii) only terms first order in derivatives in $\div S_4$ remain.
Case (i) requires $\alpha_2=1$ which already cancels the leading $1/\square$ behavior 
of the amplitude~\eqn{qAMP}. Case (ii) holds whenever $\alpha_2=1$ and $\alpha_3=0=\beta_1+6\beta_2$. By the same reasoning as above this gives again an amplitude $\sim \square+6$.
\end{enumerate}
Clearly these results are somewhat formal, though it is at least encouraging that                      the symmetric algebra technology allows their computations to be carried out efficiently.
We discuss their underlying physical principles and interpretation further in the Conclusions.

\section{Conclusions}

We have carried out an initial analysis of the interactions available to partially 
massive free gauge theories, particularly for the simplest four dimensional $s=2$ case. 
[The formal generalization to 
higher~$s$ and dimensionality is straightforward. Note, in particular that interactions
$\sim \int J^\mu D^{\nu_1}\cdots D^{\nu_{s-1}} A_{\nu_1\ldots \nu_{s-1}\mu}$ will yield 
a string of ever increasing short range interactions.]
Having motivated the choice of vector currents, primarily coupled to the model's helicity~1 excitations, rather 
than that of tensors to helicity 2, we first recast the free field into Maxwell-like form in terms of first derivative, but still gauge invariant field strengths. The coupling of the vector $J^\mu$ to the tensor field required a derivative 
coupling, and led to one-particle tree exchange amplitudes very different from the usual Maxwell 
$\int J(e^2/\square)J$ with its Coulomb $1/r$ falloff. The derivative couplings instead led to forces 
$\sim \int J(\square+2\Lambda)J$, with much steeper falloff/contact interactions that would superpose with the 
Maxwell ones if both couplings are present. We also constructed, and considered, the effects of all effective 
quartic derivative order actions maintaining scalar gauge invariance. Our analysis catalogued
these counterterms according to their effect on the divergence constraint of the 
free model. An open question is to determine whether adding such effective
terms to the Lagrangian produces ghost excitations. Imposing the only the scalar gauge 
invariance as a requirement even allows one to recover a Coulomb~$1/r$ interaction
but we are strongly suspicious that this type of correction engenders ghostlike excitations.
Therefore as it stands our proposal amounts to a candidate modification of dS electrodynamics.
By tuning the couplings $Q$ and $Q'$ we can always render it unobservable in tree physics
although much work remains to see if this is a sensible, let alone phenomenologically called
for, modification of dS quantum field theories.

We close with a warning. All computations in this paper pertain to a fixed de Sitter
background. Surely any genuine modification of de Sitter electrodynamics will require
a coupling of partially massless theories to gravity, a topic on which we currently have 
little to add.

\section*{Acknowledgments}
It is a pleasure to thank Rod Gover for useful discussions.
A.W. thanks the \'Ecole Normale Superieure, Paris
for hospitality.
This work was supported by the 
National Science Foundation under grants PHY04-01667
and PHY-01-40365.

\appendix

\section{Symmetric Tensor Algebra}

\label{symm}

Efficient computations involving symmetric tensors may be performed using
the formalism of~\cite{Hallowell:2005np}\footnote{In that work the generalization to spinors was
also given and has recently been applied to (A)dS fermionic higher spin action 
principles~\cite{Metsaev:2006zy}.}. Symmetric tensors are viewed 
as functions of {\it commuting} differentials $dx^\mu$ as suggested by the notation
for the metric tensor $ds^2=dx^\mu g_{\mu\nu}dx^\nu$. In addition the operation
$\partial_\mu\equiv d/d(dx^\mu)$ is introduced whereby
\be
[\partial_\mu, dx^\nu]=\delta^\nu_\mu\, .
\ee
To avoid confusion, note that the symbol $\partial_\mu$ does not act on functions
of the space time coordinates such as $g_{\mu\nu}(x)$ or $A_{\mu\nu}(x)$.
In this notation an $s$-index symmetric tensor is denoted
\be
\Phi=\varphi_{\mu_1\ldots\mu_s}dx^{\mu_1}\cdots dx^{\mu_s}\, ,\label{Phi}
\ee
but sums of tensors with differing number of indices are also permitted. The object $\Phi$
in~\eqn{Phi} is in fact an eigenvector of the ``index operator'' 
\be
\N\equiv dx^\mu \partial_\mu\, ,
\ee
whose job is to count indices. Component-wise, $\N:\varphi_{\mu_1\ldots\mu_s}\mapsto s \varphi_{\mu_1\ldots\mu_s}$.
 Further useful operations and their actions on component fields are
\bea
\g\equiv dx^\mu dx^\nu g_{\mu\nu}&:&\varphi_{\mu_1\ldots\mu_s}\mapsto g_{(\mu_1\mu_2} \varphi_{\mu_3\ldots\mu_{s+2})}\nn\\[2mm]
\grad\equiv dx^\mu D_\nu\ &:&\varphi_{\mu_1\ldots\mu_s}\mapsto D_{(\mu_1} \varphi_{\mu_2\ldots\mu_{s+1})}\nn\\[2mm]
\div\equiv D^\mu \partial_\mu \:\ &:&\varphi_{\mu_1\ldots\mu_s}\mapsto sD^\mu \varphi_{\mu\mu_1\ldots\mu_{s-1}}\nn\\[2mm]
\tr\equiv g^{\mu\nu} \partial_\mu \partial_\nu\ &:&\varphi_{\mu_1\ldots\mu_s}\mapsto s (s-1)\varphi^\mu{}_{\mu\mu_1\ldots\mu_{s-2}}\, .
\eea
Mnemonically: $\g$ and $\grad$ multiplies by the metric/covariant derivative and totally symmetrizes, while $\tr$ and $\div$ are the trace
and  divergence operators. The gradient operator should be viewed as the symmetric tensor generalization of the Poincar\'e exterior 
derivative $d$ for differential forms.

The advantage of these operators is their algebra\footnote{Note that  $(\g,\N,\tr)$ generate the 
$sl(2,{\mathbb R})$ Lie algebra while
$(\div,\grad)$ form its doublet representation.}
\bea
&[\N,g]=2\g\, ,\quad
[\N,\grad]=\grad\, ,\quad
[\N,\div]=-\div\, ,\quad
[\N,\tr]=-2\tr\, ,&\nn\\[3mm]
&{}[\tr,\g]=2\N+4d\, ,&
\eea
valid for any $d$-dimensional Riemannian manifold. When this manifold is flat
\be
[\div,\grad]=\Delta\, ,
\ee
and $\Delta$ is the Laplacian. In general, the commutator of $\div$ and $\grad$ equals the Laplace operator
plus somewhat complicated curvature terms. For constant curvature manifolds such as de Sitter space there
is a beautiful simplification observed long ago in a mathematical context by Lichnerowicz~\cite{Lichnerowicz}, namely
\be
[\div,\grad]=\square + 2 {\bf c} \, ,
\ee
where we employ units
\be
\Lambda=d-1\, ,
\ee 
and
\be
{\bf c}=\g\tr-\N(\N+d-2)
\ee
is the quadratic Casimir for the $sl(2,{\mathbb R})$ Lie algebra $(\g,\N,\tr)$ while 
\be
\square=\Delta+{\bf c}\, ,
\ee
is the Lichnerowicz wave operator. Importantly, it is central, {\it i.e.} commutes with all the above operations.

Finally as an example of the utility of this formalism, we spell out the partially massless spin~2 system.
Writing $A\equiv A_{\mu\nu}dx^\mu dx^\nu$, the field equations~\eqn{eom} are simply
\be
G A=0\, ,
\ee
with
\be
G=\square + 4-\grad\div+\frac12(\grad^2\tr+\g\div^2)
           -\frac12\g(\square+1)\tr\, .\label{G}
\ee
Gauge invariance and the Bianchi identity are expressed by the equalities
\be
(\div^2+\tr)G=0=G(\grad^2+\g)\, ,
\ee
which may be easily verified using the above algebra. The constraint follows because
\be
\div G = -2(\div-\grad\tr)\, .\label{Gcon}
\ee
Finally we emphasize that this algebra can be easily implemented in an algebraic manipulation 
program such as FORM~\cite{Vermaseren:2000nd} which facilitates extremely rapid computations.

\end{document}